# On the distribution of Massive White Dwarfs and Its Implication for Accretion Induced Collapse

## Ali Taani

*Applied Science Department, Aqaba University College , Al-Balqa Applied University, 19 117, Salt , Jordan*

*Physics department, University of Sharjah, P. O. Box 27272 Sharjah, United Arab Emirate.*

ali82taani@gmail.com , ali.taani@bau.edu.jo

**ABSTRACT:** A White Dwarf (WD) star and a main-sequence companion may interact through their different stellar evolution stages. This sort of binary population has historically helped us improve our understanding of binary formation and evolution scenarios. The data set used for the analysis consists of 115 well-measured WDs masses obtained by the Sloan Digital Sky Survey (SDSS). A substantial fraction of these systems could potentially evolve and reach the Chandrasekhar limit, and then undergo an Accretion Induced Collapse (AIC) to produce millisecond pulsars (MSPs). I focus my attention mainly on the massive WDs ($M_{WD} \geq 1 M_\odot$), that are able to grow further by mass-transfer phase, to reach the Chandrasekhar mass in stellar binary systems. I derive a mean value of $M \sim 1.15 \pm 0.2 M_\odot$. In the framework of the AIC process, such systems are considered to be good candidates for the production of MSPs. The implications of the results presented here to our understanding of binary MSPs evolution are discussed. As a by-product of my work, I present an updated distribution of all known pulsars in Galactic coordinates pattern.

*Subject headings:* Stars: neutron stars, white dwarfs, cataclysmic variables, x-ray binaries, fundamental parameters

## Introduction

Standard stellar evolution models predict that all main-sequence stars in our galaxy with a mass lower than $10 M_\odot$ will end their lives as white dwarfs (WD), which represent more than 97% of all stars in the galaxy. Because their structural and evolutionary properties are relatively well understood, the characteristics and the physical parameters of WD populations can be used to constrain the evolutionary models and assess the amount of mass, which was lost by their progenitor stars due to the stellar evolution process (Warner 2002; Zorotovic 2010). We only know of a few confirmed examples of very massive WDs (> 1.2 M$_\odot$). Those binary systems are of particular interest because a small amount of accreted mass could drive them above the Chandrasekhar limit, beyond which they become gravitationally unstable (Mereghetti 2009). Sloan Digital Sky Survey1 (SDSS) has provided us with a very large sample of WD/main-sequence (WDMS) binaries (Abazajian et al. 2009). This survey allowed us to make a meaningful contribution in terms of the pulsar production. One needs to implement the accretion-induced collapse (AIC) of a WD, which may lead to the formation of a proto neutron star (see: Nomoto 1987; Taani et al. 2012ab; Taani et al. 2013). This process represents a path alternative to thermonuclear disruption of accreting WDs in Type Ia supernovae (SNe). The scenario for O-Ne-Mg WD begins once the nuclear reaction starts at the center, the burning propagates throughout the entire star, and the ignition occurs in the interior. The burning that propagates outward makes central temperatures and pressures high enough to lead instead to a collapse of the WD, and induce the contraction of the star. This continues until it reaches the Chandrasekhar limit, and finally the star collapses homogeneously leading to the formation of a neutron star. Consequently, the AIC can be a significant contribution to the total number and evolution of low mass X-ray binaries (van den Heuvel 2004). In the AIC scenario, the explosion energy is expected to be small ($10^{49}$ erg s$^{-1}$) and the resulting transient short-lived, making it hard to detect by electromagnetic means (Darbha 2010) and neutrino means alone (Metzger et al. 2009). Gravitational-wave observations may provide crucial information necessary to reveal a potential AIC. On the other hand, the coalescence of double CO WDs of which the total mass exceeds the Chandrasekhar limit is also likely to end in a collapse induced by electron capture to form a rapidly rotating neutron star, rather than an SNe Ia (Podsiadlowski et al. 2004). However, the nature of the Type Ia progenitors, as well as their precise explosion mechanism, remains a subject of active investigation, both observationally as well as theoretically (Han 2008; Wang & Han 2010). In either AIC or a WD-WD merger, the WD will be rapidly rotating prior to collapse and must eject a sizable fraction of its mass (~0.1 - 0.2M$_\odot$) into a disk during the collapse in order to conserve angular momentum (Bagchi 2011, Taani et al 2012a; Taani et al. 2014).

**1. The Observed Population of WDs**

Our present study of SDSS has the following aspects. First, the large and more homogeneous sample of SDSS 2, now available, allows us to test for possible dependencies of the formation and evolution of WD on the stellar masses at the onset of common envelope evolution. Second, I select the sample of massive WDs in binary systems in the range of M ≥ 1M$_\odot$ (which is considered a relatively small number) causing a higher accretion rate onto the WD, which can reach the Chandrasekhar limit. Finally, the star collapses homogeneously leading to the formation of a rapidly rotating pulsar. Potential biases affecting the SDSS sample have been analyzed with respect to the WD mass distribution (Rebassa-Mansergas et al. 2011). The sample of WDs analyzed here consists of 115 sources (see Table 1). Fig. 1 shows the Gaussian distribution with mean at $M_{WD} \sim 1.15 M_\odot$, where the median mass of these sources is $M_{WD} = 1.181 \pm 0.16 M_\odot$. Furthermore, the knowledge of this distribution is therefore fundamental to understanding the mechanisms involved in the final stages of stellar evolution. I fit the Gauss function to the mass distributions. The Gauss function I chose reads,

$$y = y_0 + \frac{A}{w_0\sqrt{\frac{\pi}{2}}} \exp\left(-2\left(\frac{x - x_0}{w_0}\right)^2\right) \quad (1)$$

where $y_0$, $x_0$, $w_0$ and $A$ are offset of y-axis, center of x-axis, width and area represented by the curve respectively. The value of $R$ expresses the quality of the fitted result. The fitting results are listed in Table 2.

It is noteworthy to mention here that the AIC process leads to millisecond pulsar (MSP) with mass less than Chandrasekhar limit (Zhang et al. 2011). Since the binding energy also increases, this effect is significant in the estimation of the amount of mass accretion (Bagchi 2011). If matter is accreted at a rate of $\dot{M}_\odot \sim 10^{16} gs^{-1}$ (Zhang & Kojima 2006) and the total mass accreted exceeds a critical value $\Delta M_{crit} \sim 0.1 - 0.2 M_\odot$ within $10^9$ yrs, then the massive WD will be recycled to become a MSP after the mass reaches the Chandrasekhar limit. One can consider an estimate of the amount of gravitational binding energy (mass loss), due to the conversion from ONeMg WD matter to MSP. Let us begin with the gravitational energy in NS (Shapiro & Teukolsky 1983)

$$E = \frac{GM^2}{R} \qquad 2$$

Where $G$ is the gravitational constant, M the mass companion, R the NS radius

$$E = \left(\frac{2GM}{c^2}\right)\frac{Mc^2}{2R} \rightarrow \frac{R_S M c^2}{2R} \qquad 3$$

where $R_S$ is Schwartzchild radius ($R_S = 2GM/c^2 \sim 3$ km for ($M=1M_\odot$))

$$E = \eta M c^2 \qquad 4$$

where $\eta = R_S/2R$, thus

$$E \sim 0.1 M c^2 \qquad 5$$

As a result, around ~ 10% of mass will goes as binding energy and reduce the mass less than 1.4M⊙ with about 0.14M☉, where the total energy is

$$Mc^2(1-\eta) \qquad 6$$

This effect is significant in the estimation of the amount of mass accretion. However, unfortunately, most of the studies argued that the binding energy of the NS is not commonly considered (Bagchi 2011). More modeling is required to support such a conclusion.

Fig. 2 shows the spatial density distribution of the current sample of all known WD systems throughout the Galaxy, the scale-height of the WDs is somewhat completed and uniformly distributed.

## 2. Summary and Conclusions

Interacting binaries can play an important role in terms of studying the mass transfer and orbital motion of the companions as well as the effects of binary evolution on the final fate of the compact objects and SNe.

The main conclusion of the present study is that the evolution of accreting WDs is seriously affected on the pulsar production via the mass and angular momentum accretion during the AIC scenario, which requires substantial adjustments to the current picture on MSP progenitor population.
(* not clear, cannot be directly concluded from the above treatment)

In the frame of this process, the AIC scenario must be invoked to produce binary MSPs, since it is expected to turn into normal magnetized rotating neutron stars, which in binaries can evolve into MSPs through the recycling process. In addition, it could provide us with a better match to the period distributions of some types of binary MSPs. One example is the double relativistic pulsar PSR J0737–3039 (Burgay et al. 2003; Lyne et al. 2004), and it seems to be an interesting issue requiring further studies for the analysis of the formation of neutron stars in Galactic disk as well as globular clusters.

Finally, it is worth mentioning here that the most highly magnetized WDs are massive as well as isolated. This depends on the strength of the WDs magnetic field. The matter flowing through the inner Lagrangian point (L1) can form either a full or a partial accretion disk (Zhang et al. 2009), or else follow the magnetic field lines down to the surface at the magnetic poles. Thus it is important to monitor long orbital period of WD systems in order to gain insight into the dynamics and binary evolution of those systems that most likely to undergo the AIC process. Future work can go further using the model by Zhang et al. (2009) to estimate the bottom magnetic field for the observed sample of WD.

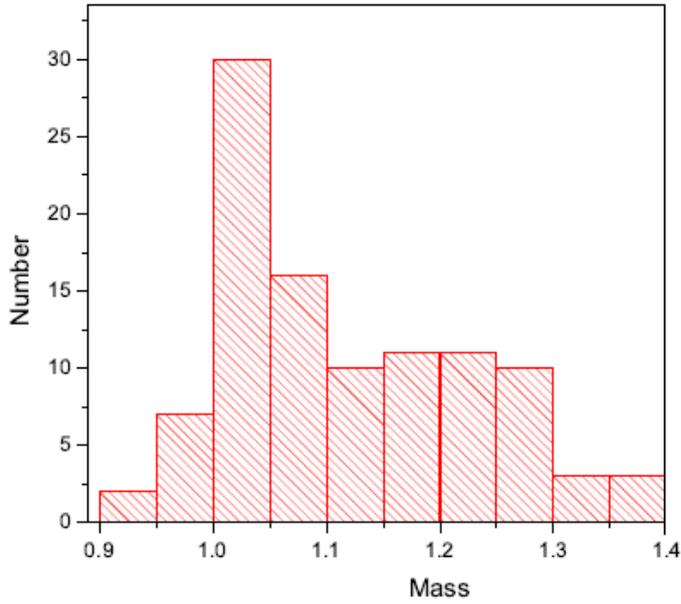

Fig. 1.— Histograms of the mass distribution of WDs. The distribution is relatively Gaussian.

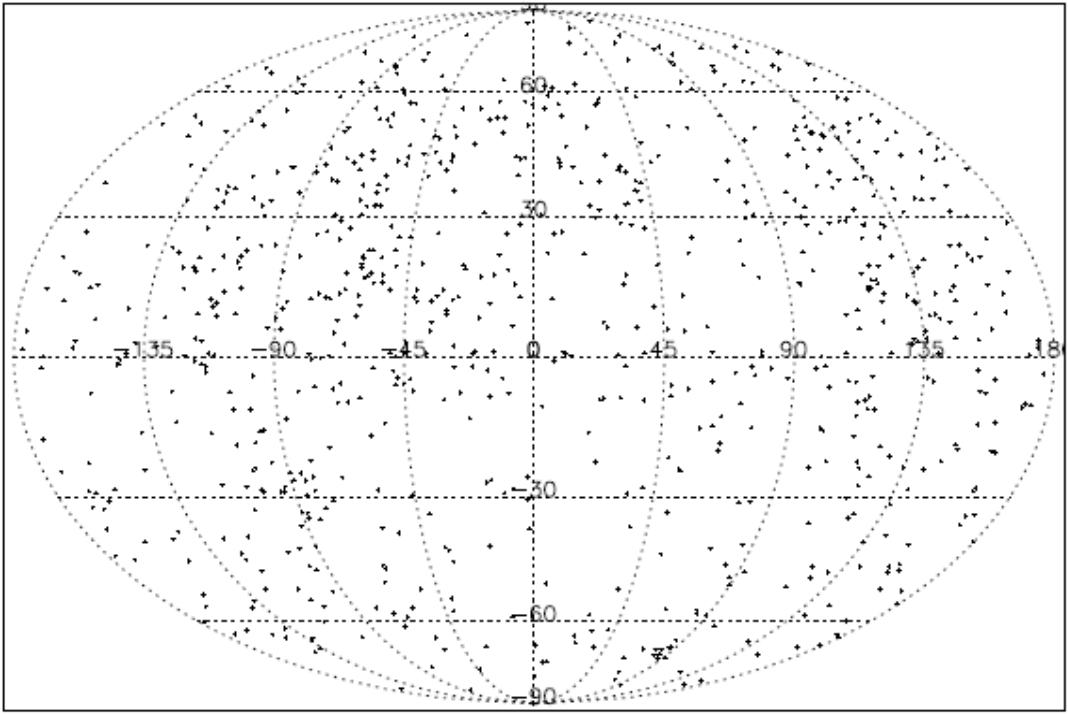

Fig2. The full-sky distribution of all known WDs, shown in Galactic coordinates in which the plane of the Galaxy is at the equator, and the Galactic center is at the origin. The data are taken from Ritter & Kolb (2014).

**Table (1): Parameters of binary systems of WDs with compact companions. Data were taken from SDSS catalogue**

| IAU Name | $M_{wd}(M_\odot)$ | Errors $M_{wd}(M_\odot)$ | $M_{seccandary}(M_\odot)$ | $E_{rros}M_{sec}(M_\odot)$ | $P_{orb}(d)$ | $R_{wd}(R_\odot)$ |
|---|---|---|---|---|---|---|
| SDSSJ001247.18+001048.7 | 1.23 | 0.442 | 0.38 | 0.072 | 0 | 0.00538 |
| SDSSJ001855.19+002134.5 | 1.24 | 0.391 | 0.38 | 0.072 | 0 | 0.00526 |
| SDSSJ002157.90-110331.6 | 1.08 | 0.051 | 0.319 | 0.09 | 0 | 0.0071 |
| SDSSJ002959.94+001132.7 | 1.15 | 0.386 | 0.431 | 0.1 | 0 | 0.00637 |
| SDSSJ003301.53+005716.9 | 1.19 | 0.286 | 0.38 | 0.072 | 0 | 0.00588 |
| SDSSJ003643.33+252754.9 | 1 | 0.113 | 0.118 | 0.004 | 0 | 0.00807 |
| SDSSJ003731.07+010947.0 | 0.98 | 0.629 | 0.319 | 0.09 | 0 | 0.00826 |
| SDSSJ004511.95+250330.9 | 1.13 | 0.037 | 0 | 0 | 0 | 0.00654 |
| SDSSJ004604.14+011037.4 | 1.16 | 0.386 | 0.464 | 0.088 | 0 | 0.00618 |
| SDSSJ005245.11-005337.2 | 1.26 | 0.365 | 0.319 | 0.09 | 2.735 | 0.00502 |
| SDSSJ011634.11+002956.5 | 0.93 | 0.348 | 0.319 | 0.09 | 0 | 0.00898 |
| SDSSJ013504.31-085919.0 | 1.23 | 0.222 | 0.319 | 0.09 | 0 | 0.00538 |
| SDSSJ015434.31-010611.1 | 0.98 | 0.688 | 0.255 | 0.124 | 0 | 0.00818 |
| SDSSJ015657.37-003341.6 | 0.91 | 0.357 | 0.464 | 0.088 | 0 | 0.0093 |
| SDSSJ020925.74+064213.9 | 1.15 | 0.15 | 0.319 | 0.09 | 0 | 0.00644 |
| SDSSJ021309.19-005025.4 | 1.02 | 0.423 | 0.38 | 0.072 | 0 | 0.00779 |
| SDSSJ024642.55+004137.2 | 0.95 | 0.099 | 0.38 | 0.072 | 17.43 | 0.00873 |
| SDSSJ024645.89-010624.1 | 1.31 | 0.355 | 0.38 | 0.072 | 0 | 0.0044 |
| SDSSJ030351.97+003548.4 | 1.307 | 0.281 | 0.431 | 0.1 | 0 | 0.00448 |
| SDSSJ030904.82-010100.8 | 0.977 | 0.406 | 0.355 | 0.072 | 0 | 0.00846 |
| SDSSJ031206.82-002145.4 | 1.13 | 0.651 | 0.319 | 0.09 | 0 | 0.00654 |
| SDSSJ032656.44+002232.1 | 1.105 | 0.447 | 0.431 | 0.1 | 0 | 0.00694 |
| SDSSJ033804.40+002740.3 | 1.37 | 0.091 | 0 | 0 | 0 | 0.0038 |
| SDSSJ035912.46-044630.2 | 1.24 | 1.058 | 0.38 | 0.072 | 0 | 0.00525 |
| SDSSJ052624.54+621344.2 | 1.07 | 0.037 | 0.38 | 0.072 | 0 | 0.00723 |
| SDSSJ073250.34+393633.9 | 0.905 | 0.074 | 0 | 0 | 0 | 0.00917 |
| SDSSJ074207.89+275845.1 | 1.385 | 0.054 | 0.319 | 0.09 | 0 | 0.00355 |
| SDSSJ075325.93+164132.7 | 1.19 | 0.411 | 0.464 | 0.088 | 0 | 0.00581 |
| SDSSJ080304.61+121810.3 | 1.04 | 0.274 | 0.431 | 0.1 | 0 | 0.00768 |
| SDSSJ080653.95+160729.8 | 0.91 | 0.269 | 0.38 | 0.072 | 0 | 0.0091 |
| SDSSJ081312.09+324758.6 | 0.9 | 0.204 | 0.255 | 0.124 | 0 | 0.00917 |
| SDSSJ081811.70+173224.5 | 1.04 | 0.173 | 0.38 | 0.072 | 0 | 0.00768 |
| SDSSJ081831.07-010923.1 | 1.14 | 0.311 | 0.38 | 0.072 | 0 | 0.00648 |
| SDSSJ083722.44+265417.3 | 0.94 | 0.065 | 0.319 | 0.09 | 0 | 0.00877 |
| SDSSJ083827.09+415015.5 | 1.07 | 0.365 | 0.464 | 0.088 | 0 | 0.00723 |
| SDSSJ083920.48+125959.5 | 1.12 | 0.154 | 0.118 | 0.004 | 0 | 0.0066 |
| SDSSJ084009.24+281201.9 | 1.07 | 0.262 | 0.38 | 0.072 | 0 | 0.00756 |
| SDSSJ084056.91+275513.7 | 1.07 | 0.203 | 0.38 | 0.072 | 0 | 0.0073 |
| SDSSJ084307.27+122610.1 | 0.95 | 0.129 | 0.196 | 0.085 | 0 | 0.0086 |
| SDSSJ084400.82+052305.7 | 0.94 | 0.284 | 0.319 | 0.09 | 0 | 0.00893 |
| SDSSJ085847.47+371115.5 | 1.03 | 0.366 | 0.464 | 0.088 | 0 | 0.00849 |
| SDSSJ091143.09+222748.8 | 1.46 | 0.051 | 0.38 | 0.072 | 0 | 0.0027 |
| SDSSJ091309.70+223346.7 | 0.98 | 0.167 | 0.431 | 0.1 | 0 | 0.00828 |
| SDSSJ092200.71+181714.1 | 1.02 | 0.187 | 0.38 | 0.072 | 0 | 0.0078 |
| SDSSJ092313.99+205119.9 | 1.16 | 0.18 | 0.38 | 0.072 | 0 | 0.00623 |
| SDSSJ092433.98+204020.0 | 1.04 | 0.113 | 0.319 | 0.09 | 0 | 0.00754 |
| SDSSJ093236.83+053026.6 | 1.02 | 0.16 | 0.464 | 0.088 | 0 | 0.0078 |
| SDSSJ093426.60+053753.6 | 0.93 | 0.277 | 0.319 | 0.09 | 0 | 0.00912 |

| Name | | | | | | |
|---|---|---|---|---|---|---|
| SDSSJ093427.91+204658.6 | 1.03 | 0.418 | 0.196 | 0.085 | 0 | 0.00759 |
| SDSSJ093526.43+245423.4 | 1.02 | 0.224 | 0.431 | 0.1 | 0 | 0.00785 |
| SDSSJ093632.33+341932.6 | 1.03 | 0.114 | 0.319 | 0.09 | 0 | 0.00767 |
| SDSSJ094002.40+534202.9 | 0.95 | 0.159 | 0.255 | 0.124 | 0 | 0.00868 |
| SDSSJ094542.61+173859.9 | 1.01 | 0.185 | 0.472 | 0.062 | 0 | 0.00821 |
| SDSSJ094720.94+111734.7 | 1.12 | 0.193 | 0.464 | 0.088 | 0 | 0.0074 |
| SDSSJ094821.30+365935.3 | 1.26 | 0.135 | 0.38 | 0.072 | 0 | 0.00501 |
| SDSSJ094952.73+012603.4 | 1.14 | 0.248 | 0.319 | 0.09 | 0 | 0.00648 |
| SDSSJ100015.18+304330.5 | 0.93 | 0.214 | 0.255 | 0.124 | 0 | 0.00898 |
| SDSSJ100609.18+004417.0 | 0.93 | 0.083 | 0.118 | 0.004 | 6.7313 | 0.00879 |
| SDSSJ101006.92+301211.3 | 1.2 | 0.158 | 0.255 | 0.124 | 0 | 0.00568 |
| SDSSJ101116.29+182749.5 | 0.92 | 0.229 | 0 | 0 | 0 | 0.00992 |
| SDSSJ101124.31+644655.5 | 1.15 | 0.166 | 0.118 | 0.004 | 0 | 0.00636 |
| SDSSJ101614.69+490930.3 | 1.01 | 0.168 | 0.319 | 0.09 | 0 | 0.00801 |
| SDSSJ102213.46+294119.9 | 1.275 | 0.315 | 0.319 | 0.09 | 0 | 0.00483 |
| SDSSJ104459.32+360554.7 | 1.24 | 0.052 | 0.255 | 0.124 | 0 | 0.00519 |
| SDSSJ105038.62+413834.5 | 0.95 | 0.085 | 0.196 | 0.085 | 0 | 0.00861 |
| SDSSJ110736.88+612232.8 | 1 | 0.043 | 0.464 | 0.088 | 0 | 0.00815 |
| SDSSJ114720.00+112813.3 | 0.91 | 0.296 | 0.464 | 0.088 | 0 | 0.00909 |
| SDSSJ114913.52-014728.6 | 0.9 | 0.057 | 0.319 | 0.09 | 0 | 0.00917 |
| SDSSJ120222.10+411810.8 | 0.97 | 0.414 | 0.38 | 0.072 | 0 | 0.0087 |
| SDSSJ121928.05+161158.7 | 0.98 | 0.124 | 0.196 | 0.085 | 0 | 0.00818 |
| SDSSJ122634.97+322020.8 | 1.36 | 0.108 | 0.319 | 0.09 | 0 | 0.00379 |
| SDSSJ122850.46-022509.4 | 1.18 | 0.379 | 0 | 0 | 0 | 0.00599 |
| SDSSJ122930.65+263050.4 | 1.04 | 0.077 | 0.38 | 0.072 | 0 | 0.0077 |
| SDSSJ125207.48+444827.8 | 0.9 | 0.231 | 0 | 0 | 0 | 0.00938 |
| SDSSJ125919.51+321935.4 | 0.9 | 0.116 | 0.38 | 0.072 | 0 | 0.00915 |
| SDSSJ131156.69+544455.8 | 1.19 | 0.036 | 0.431 | 0.1 | 0 | 0.0058 |
| SDSSJ132652.62+000855.3 | 0.97 | 0.446 | 0.464 | 0.088 | 0 | 0.00863 |
| SDSSJ134557.68+330050.6 | 0.97 | 0.097 | 0.464 | 0.088 | 0 | 0.00839 |
| SDSSJ134714.30+412909.6 | 1.25 | 0.142 | 0.431 | 0.1 | 0 | 0.00513 |
| SDSSJ135502.75+574058.3 | 1.46 | 0.049 | 0 | 0 | 0 | 0.0027 |
| SDSSJ142241.90+513537.9 | 1.23 | 0.043 | 0 | 0 | 0 | 0.00537 |
| SDSSJ143519.16+362952.0 | 1.02 | 0.121 | 0.319 | 0.09 | 0 | 0.00773 |
| SDSSJ145514.56-022822.2 | 0.94 | 0.381 | 0.149 | 0.075 | 0 | 0.00866 |
| SDSSJ151714.96+423924.7 | 1.25 | 0.02 | 0.38 | 0.072 | 0 | 0.00507 |
| SDSSJ151817.97+455334.1 | 0.97 | 0.245 | 0.464 | 0.088 | 0 | 0.00833 |
| SDSSJ154609.98+200320.6 | 1.12 | 0.208 | 0.464 | 0.088 | 0 | 0.00674 |
| SDSSJ154928.56+385419.3 | 0.93 | 0.222 | 0.431 | 0.1 | 0 | 0.00882 |
| SDSSJ155332.86+045735.1 | 0.91 | 0.174 | 0.38 | 0.072 | 0 | 0.0091 |
| SDSSJ155933.42+340502.5 | 0.96 | 0.254 | 0 | 0 | 0 | 0.00838 |
| SDSSJ160821.47+085149.9 | 0.98 | 0.083 | 0.196 | 0.085 | 9.936 | 0.00826 |
| SDSSJ160824.57+285524.9 | 1.06 | 0.286 | 0.319 | 0.09 | 0 | 0.00742 |
| SDSSJ161715.27+081849.0 | 1.02 | 0.295 | 0.464 | 0.088 | 0 | 0.0079 |
| SDSSJ161726.96+385557.5 | 1.02 | 0.282 | 0.319 | 0.09 | 0 | 0.00807 |
| SDSSJ162324.05+343647.7 | 0.97 | 0.071 | 0.118 | 0.004 | 0 | 0.00832 |
| SDSSJ170127.36+253302.6 | 1.01 | 0.08 | 0.38 | 0.072 | 0 | 0.00793 |
| SDSSJ170843.52+215829.0 | 1.27 | 0.224 | 0.464 | 0.088 | 0 | 0.00502 |
| SDSSJ171145.42+555444.4 | 0.94 | 0.04 | 0.319 | 0.09 | 0 | 0.01017 |
| SDSSJ171411.14+294508.2 | 0.91 | 0.231 | 0.319 | 0.09 | 0 | 0.00933 |
| SDSSJ172008.58+565211.8 | 1.29 | 0.276 | 0.464 | 0.088 | 0 | 0.00471 |
| SDSSJ205059.37-000254.3 | 0.95 | 0.099 | 0.319 | 0.09 | 0 | 0.00862 |
| SDSSJ205316.52-010616.5 | 0.9 | 0.112 | 0.38 | 0.072 | 0 | 0.00915 |
| SDSSJ210624.12+004030.2 | 1.03 | 0.105 | 0.464 | 0.088 | 0 | 0.00774 |

| Name | | | | | | |
|---|---|---|---|---|---|---|
| SDSSJ210751.44+005854.4 | 1.01 | 0.268 | 0 | 0 | 0 | 0.00811 |
| SDSSJ211132.76+011522.2 | 0.96 | 0.124 | 0.431 | 0.1 | 0 | 0.00847 |
| SDSSJ211205.31+101427.9 | 1.06 | 0.051 | 0.196 | 0.085 | 2.2152 | 0.00744 |
| SDSSJ214447.51+004201.5 | 1.01 | 0.446 | 0.149 | 0.075 | 0 | 0.00785 |
| SDSSJ215744.77-004015.1 | 1.2 | 0.345 | 0.319 | 0.09 | 0 | 0.00575 |
| SDSSJ220848.99+122144.7 | 1.24 | 0.134 | 0 | 0 | 45.67 | 0.00568 |
| SDSSJ224522.42-000109.5 | 1.11 | 0.222 | 0.38 | 0.072 | 0 | 0.00686 |
| SDSSJ224932.02+000645.7 | 1.08 | 0.217 | 0.196 | 0.085 | 0 | 0.00709 |
| SDSSJ231014.62+001439.9 | 1.07 | 0.32 | 0.38 | 0.072 | 0 | 0.00722 |
| SDSSJ232527.81-005416.7 | 0.93 | 0.194 | 0.255 | 0.124 | 0 | 0.00898 |
| SDSSJ232624.72-011327.2 | 1.46 | 0.372 | 0.38 | 0.072 | 0 | 0.00269 |
| SDSSJ232816.06+010036.0 | 1.31 | 0.931 | 0.38 | 0.072 | 0 | 0.00447 |
| SDSSJ234749.84+431424.6 | 0.99 | 0.079 | 0.255 | 0.124 | 0 | 0.0082 |
| SDSSJ235324.74+351623.2 | 1.08 | 0.211 | 0.255 | 0.124 | 0 | 0.00709 |
| SDSSJ011355.84-093938.0 | 1.24 | 0.156 | 0.319 | 0.09 | 0 | 0.00519 |
| SDSSJ041518.90+165238.2 | 1.28 | 0.066 | 0 | 0 | 0 | 0.00483 |
| SDSSJ042437.67+063408.2 | 0.92 | 0.169 | 0.464 | 0.088 | 0 | 0.0099 |
| SDSSJ044542.27+120246.7 | 1.15 | 0.168 | 0.255 | 0.124 | 0 | 0.00635 |
| SDSSJ064411.89+285301.1 | 0.96 | 0.122 | 0 | 0 | 0 | 0.00936 |
| SDSSJ064715.54+275948.3 | 1 | 0.13 | 0 | 0 | 0 | 0.00799 |
| SDSSJ085224.02+111520.8 | 0.93 | 0.407 | 0.319 | 0.09 | 0 | 0.00898 |
| SDSSJ093349.93+151718.5 | 1 | 0.125 | 0.431 | 0.1 | 0 | 0.00825 |
| SDSSJ100811.87+162450.4 | 1.23 | 0.07 | 0.464 | 0.088 | 0 | 0.00575 |
| SDSSJ113223.69+225313.1 | 0.98 | 0.053 | 0.464 | 0.088 | 0 | 0.00845 |
| SDSSJ113511.13+000923.9 | 0.935 | 0.157 | 0.319 | 0.09 | 0 | 0.00875 |
| SDSSJ125645.47+252241.6 | 1.01 | 0.06 | 0.255 | 0.124 | 0 | 0.00818 |
| SDSSJ140516.05+232246.9 | 0.94 | 0.069 | 0 | 0 | 0 | 0.00883 |
| SDSSJ141451.60+193638.9 | 1.07 | 0.13 | 0.196 | 0.085 | 0 | 0.00716 |
| SDSSJ173430.11+335407.5 | 1.19 | 0.133 | 0.319 | 0.09 | 0 | 0.00586 |
| SDSSJ213225.96+001430.5 | 0.92 | 0.125 | 0.319 | 0.09 | 0 | 0.00904 |
| SDSSJ011123.90+000935.2 | 0.93 | 0.202 | 0.431 | 0.1 | 0 | 0.00899 |
| SDSSJ014232.59-083528.4 | 1.07 | 0.05 | 0.38 | 0.072 | 0 | 0.00722 |
| SDSSJ025347.51+335221.0 | 1.02 | 0.121 | 0.431 | 0.1 | 0 | 0.00808 |
| SDSSJ044831.02+214909.8 | 1.01 | 0.11 | 0.38 | 0.072 | 0 | 0.00793 |
| SDSSJ072130.60+374228.3 | 0.96 | 0.102 | 0.319 | 0.09 | 0 | 0.00854 |
| SDSSJ072434.72+321609.4 | 0.98 | 0.051 | 0.319 | 0.09 | 0 | 0.00835 |
| SDSSJ085223.75+071326.0 | 0.92 | 0.231 | 0.319 | 0.09 | 0 | 0.00896 |
| SDSSJ102102.25+174439.9 | 1.06 | 0.087 | 0.319 | 0.09 | 0 | 0.00768 |
| SDSSJ112308.40-115559.3 | 1.26 | 0.07 | 0.255 | 0.124 | 0 | 0.00501 |
| SDSSJ112651.03-081640.1 | 1.26 | 0.125 | 0.431 | 0.1 | 0 | 0.00501 |
| SDSSJ120953.67+185815.7 | 0.96 | 0.126 | 0.464 | 0.088 | 0 | 0.00883 |
| SDSSJ130012.49+190857.4 | 1.09 | 0.103 | 0.319 | 0.09 | 7.391 | 0.0069 |
| SDSSJ135207.77+185033.8 | 1.1 | 0.122 | 0.38 | 0.072 | 0 | 0.00684 |
| SDSSJ141052.79+375435.6 | 1.03 | 0.074 | 0.464 | 0.088 | 0 | 0.00767 |
| SDSSJ142503.62+073846.4 | 0.97 | 0.13 | 0.38 | 0.072 | 0 | 0.0087 |
| SDSSJ142631.93+091621.1 | 1.4 | 0.075 | 0.38 | 0.072 | 0 | 0.00343 |
| SDSSJ142951.19+575949.0 | 1.07 | 0.131 | 0.38 | 0.072 | 13.08486 | 0.0073 |
| SDSSJ143143.83+565728.2 | 1 | 0.285 | 0.255 | 0.124 | 0 | 0.00799 |
| SDSSJ145305.77+001048.2 | 0.95 | 0.054 | 0.319 | 0.09 | 0 | 0.00861 |
| SDSSJ153009.49+384439.8 | 0.92 | 0.282 | 0.431 | 0.1 | 0 | 0.00895 |
| SDSSJ155808.49+264225.7 | 1.06 | 0.309 | 0.319 | 0.09 | 15.9016 | 0.00736 |
| SDSSJ162354.45+630640.4 | 1 | 0.089 | 0.319 | 0.09 | 53.5624 | 0.00799 |
| SDSSJ173849.76+635042.0 | 0.93 | 0.123 | 0.38 | 0.072 | 0 | 0.00898 |

Table 2 : the fitting results of the distribution for mass of WDs.

| quantity | $y_0$ | $x_c$ | $w$ | $A$ | $\sigma$ | $\chi^2$/Dof | $R^2$ |
|---|---|---|---|---|---|---|---|
| $M_{WDs}$ | 0.53 ± 0.64 | 0.79 ± 0.012 | 0.40 ± 0.03 | 12.30 ± 0.92 | 0.20 | 3.311 | 0.955 |